\documentclass[conference,letterpaper]{IEEEtran}

\usepackage{dsfont}
\usepackage{subcaption}
\usepackage{moreverb}
\usepackage{epsfig}
\usepackage{amsmath,amssymb,amsthm,mathrsfs,amsfonts,dsfont}
\usepackage{adjustbox,lipsum}
\usepackage{algorithm,algorithmic}
\usepackage{amsfonts}
\usepackage{epsfig}
\usepackage{amssymb}
\usepackage{amsmath}
\usepackage{amsthm}
\usepackage{multirow}
\usepackage{rotating}
\usepackage{graphicx}
\usepackage{tabularx}
\usepackage{array}
\usepackage{anyfontsize}
\usepackage{color,soul}
\usepackage{graphicx,dblfloatfix}
\usepackage{epstopdf}
\usepackage{blindtext}
\usepackage{amsmath}
\usepackage{amsthm,amssymb,amsmath,bm}
\usepackage{amsfonts}
\usepackage{epsfig}
\usepackage{amssymb}
\usepackage{amsmath}
\usepackage{cite}
\usepackage{graphicx}
\usepackage{fancyhdr}
\usepackage[font={small}]{caption}
\usepackage{tabularx}
\usepackage{tcolorbox}
\usepackage{cite}
\usepackage{algorithm}
\usepackage{algorithmic}
\usepackage{soul}

\DeclareMathOperator*{\argmin}{arg\,min}

\newtheorem{remark}{Remark}

\begin{document}

	\title{Age-Aware Status Update Control for Energy Harvesting IoT Sensors via Reinforcement Learning}
	
	\author{\IEEEauthorblockN{Mohammad Hatami$^1$, Mojtaba Jahandideh$^1$, Markus Leinonen$^1$, and Marian Codreanu$^2$}
		\IEEEauthorblockA{$^{1}$Centre for Wireless Communications, University of Oulu, Finland \\
			$^{2}$Department of Science and Technology, Link\"{o}ping University, Sweden \\
			Email:  mohammad.hatami@oulu\hspace{0.125em}.fi, mojtaba.jahandideh@oulu\hspace{0.125em}.fi, markus.leinonen@oulu\hspace{0.125em}.fi, marian.codreanu@liu\hspace{0.125em}.se\\
	}}

	\maketitle
	
	\begin{abstract}
		We consider an IoT sensing network with multiple users, multiple energy harvesting sensors, and a wireless edge node acting as a gateway between the users and sensors. The users request for updates about the value of physical processes, each of which is measured by one sensor. The edge node has a cache storage that stores the most recently received measurements from each sensor. Upon receiving a request, the edge node can either command the corresponding sensor to send a status update, or use the data in the cache. We aim to find the best action of the edge node to minimize the average long-term cost which trade-offs between the age of information and energy consumption. \textcolor{black}{We propose a practical reinforcement learning approach that finds an optimal policy without knowing the exact battery levels of the sensors.} 
		Simulation results show that the proposed method significantly reduces the average cost compared to several baseline methods.
	\end{abstract}
	
	
	\section{Introduction}
	Internet of Things (IoT) is a new technology which uses minimal human intervention and connects different devices and applications. IoT enables us to effectively interact with the physical surrounding environment and empower context-aware applications like smart cities \cite{Xu2014IoTSurvey}. A \textcolor{black}{typical} IoT sensing network consists of multiple wireless sensors which measure a physical quantity and communicate the measurements to a \textcolor{black}{destination} for further processing. Two special features of these networks are: 1) stringent energy limitations of battery-powered sensors which may be counteracted by \textit{harvesting} energy from environmental sources like sun, heat, and RF ambient \cite{AmbientRFEH2014},
	and 2) \textit{transient} nature of data, i.e., the sensors' measurements become outdated after a while.
	Thus, it is crucial to design IoT sensing techniques where the sensors sample and send minimal number of measurements to prolong their lifetime while providing the end users highly fresh data for time-sensitive IoT applications. The freshness of information \textcolor{black}{from the users' perspective} can be quantified by the recently emerged metric, the \textit{age of information} (AoI) \cite{AoI_Orginal_12,yates19AoI,costa16AoI}.
	
	
	We consider an IoT sensing network consisting of multiple users, multiple energy harvesting IoT sensors, and a wireless edge node. The users send requests for the physical \textcolor{black}{processes}, each of which is measured by one sensor. The edge node, which acts as a gateway between the users and the sensors, has a cache storage which stores the most recently received measurements of each physical quantity. Upon receiving a request, the edge node can either command the corresponding sensor to \textcolor{black}{sample and send a new measurement}, or use the available data in the cache. The former leads to having a fresh measurement, yet at the cost of increased energy consumption. Since the latter prevents the activation of the sensors for every single request, the sensors can stay longer in a sleep mode to save a considerable amount of energy \cite{niyato2016novel}, but the data forwarded to the users becomes stale. \textcolor{black}{This results in an inherent \textit{trade-off} between the AoI of sensing data and sensors' energy consumption.}
	
	\textbf{Contributions:} The main objective of this paper is to find the best action of the edge node at each time slot, which is called an optimal policy, to strike a balance between the AoI and energy consumption in the considered IoT sensing network. We address a realistic scenario where the edge node does not know the exact battery level of each energy harvesting sensor at each time slot, but only the level from a sensor's last update. We model the problem of finding an optimal policy as a Markov decision process (MDP). We propose  a reinforcement learning (RL) based algorithm to obtain an optimal policy that minimizes a cost function that trade-offs the AoI and energy consumption.
	Simulation results show that the proposed method significantly reduces the average cost compared to several baseline methods.

	\textbf{Related works:} 
	\textcolor{black}{RL is an \textit{online} machine learning method which learns an optimal policy through the interactions between the agent (the edge node in our case) and the environment.}
	A comprehensive survey of RL based methods for autonomous IoT networks is presented in \cite{lei2019deep}.
	In \cite{asilomar2019_caching,Sadeghi_2018_scalable}, the authors used RL to find an optimal caching policy for non-transient data (e.g., multimedia files). 
	\textcolor{black}{In \cite{abd2019reinforcement}, deep RL was used to minimize AoI in a real-time multi-node monitoring system, in which the sensors are powered through wireless energy transfer by the destination.}
	The authors in \cite{zhu2018caching} used deep RL to solve a cache replacement problem with a limited cache size and transient data in an IoT network. \textcolor{black}{Different from \cite{zhu2018caching},  we consider both energy harvesting and energy limitation of the IoT sensors.} The authors in \cite{niyato2016novel} considered \textcolor{black}{a known energy harvesting model} and proposed a threshold adaptation algorithm to maximize the hit rate in an IoT sensing network. Compared to \cite{niyato2016novel}, we include data freshness/AoI and, by assuming that the energy harvesting model is unknown, use RL to search for the optimal policy.
	
	\section{System Model and Problem Formulation}
	\subsection{Network Model}\label{sec_systemmodel}
\begin{figure}[t!]
		\centering
		\includegraphics[width=.98\columnwidth]{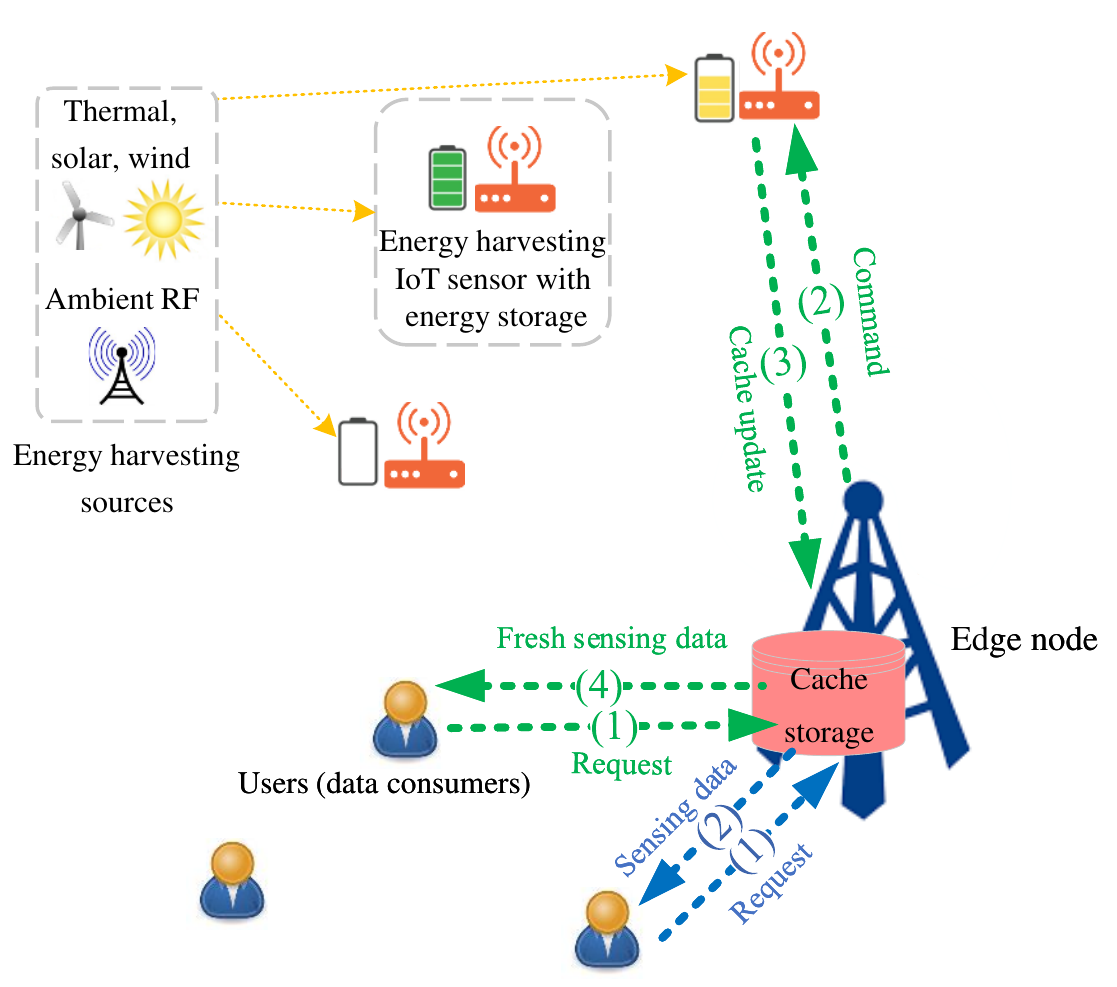}
		\caption{IoT sensing network consisting of multiple users (\textit{data consumers}), one wireless edge node (i.e., the gateway), and a set of $K$ energy harvesting wireless IoT sensors (\textit{data producers}).  \textcolor{black}{The procedure of serving a request by using fresh data is shown by green lines, and the blue lines show the procedure of serving a request by using the previous measurements already existing in the cache.}}
		\label{fig_systemmodel}
		\vspace{-0.5cm}
	\end{figure}

	We consider an IoT sensing network consisting of multiple users (\textit{data consumers}), a wireless edge node, and a set of $K$ energy harvesting sensors (\textit{data producers}), as depicted in Fig.~\ref{fig_systemmodel}. Sensor $k\in\mathcal{K}=\left\lbrace 1,\dots,K \right\rbrace$ measures independently a specific physical quantity  $f_k$, e.g., temperature or humidity. The system operates in a slotted fashion, i.e., time is divided into slots which are labeled with a discrete index ${t \in \mathbb{N}}$.
	
	We assume that there is no direct link between the users and the sensors, i.e., the edge node acts as a gateway between them. 
	\textcolor{black}{Users request for the values of physical
		quantities so that at each time slot, there can be multiple requests arriving at the edge node. We assume that the requests for the value of  physical quantities come at the beginning of each slot and the edge node sends  values to the users at the end of the same slot. Let $r_k(t) \in \{0,1\}$, $t=1,2,\dots$, denote the random process of requesting the value of $f_k$ at the beginning of slot $t$; $r_k(t) = 1$ if the value of $f_k$ is requested and  $r_k(t)=0$, otherwise.}
	
	The edge node is equipped with a cache storage that stores the most recently received measurement of each physical quantity. Upon receiving a request for the value of $f_k$ at slot $t$ (i.e., $r_k(t)=1$), the edge node can either command sensor $k$ to perform a new measurement and send a \textit{status update}, or use the previous measurement in the local cache, to serve the request.
	Let $a_k(t) \in \{0,1\}$ denote the \textit{command} action of the edge node at slot $t$; $a_k(t)=1$ if the edge node commands sensor $k$ to send a status update  and $a_k(t)=0$ otherwise.
	
	\subsection{Energy Harvesting Model}
	Sensors rely on the energy harvested from the environment. Sensor $k$ stores the harvested energy in a battery of finite size $B_k$ (units of energy).
	\textcolor{black}{For defining the cost of transmitting a status update from each sensor to the edge node, we consider the common assumption  (see e.g., \cite{bacinoglu2015age_oneunitenergy,arafa2019timely,wu2017optimal_oneunitenergy,arafa2019age,michelusi2013transmission}) that
		this transmission consumes one unit of energy\footnote{\textcolor{black}{While simple, this model encompasses the crucial energy cost of low-power sensors and thus, gives rise to the fundamental trade-off between the freshness of measurements and energy consumption of the sensors in our considered status update control problem (see Section \ref{sec_cost}). Consideration of more realistic wireless channels is an interesting future study.}}.}
	Let random variable $d_k(t) \in \left\lbrace 0 ,1\right\rbrace$ denote the action of sensor $k$ at slot $t$; $d_k(t)=1$ if sensor $k$ sends a status update to the edge node and $d_k(t)=0$ otherwise. 
	Note that  $d_k(t)$ and $a_k(t)$ can be different which is discussed in Section~\ref{subsec:partialknowledge}.

	Let $b_k(t)$ denote the battery level of sensor $k$ at the beginning of slot $t$. The evolution of the battery level of sensor $k$ can be expressed as
	\begin{equation}\label{battery_evo}
	b_k(t+1) = \min\left\lbrace  b_k(t)+e_k(t)-d_k(t) , B_k \right\rbrace,
	\end{equation}
	where $e_k(t) \in \left\lbrace 0 ,1\right\rbrace $, $t = 1,2,\dots$, is the \textit{energy arrival process} of sensor $k$. We assume that the energy arrival processes are independent and unknown to the edge node. Moreover, the energy harvested during slot $t$ can be used only in a later slot.
	
	\subsection{Status Update with Partial Knowledge of the Battery Levels}\label{subsec:partialknowledge}
	We consider a realistic environment in which the edge node is informed about the battery levels of the sensors only via the \textit{status update packets}.
	Each status update packet consists of the value of $f_k$, the generation timestamp, and the battery level of sensor $k$. \textcolor{black}{Let variable $\tilde{b}_k(t)$ denote the battery level of sensor $k$ at the beginning of that time slot in which the most recent  status update of sensor $k$ was received by the edge node.} Thus, the edge node does not know the exact battery level of the sensors at each time slot, but it only has the \textit{partial} knowledge, i.e., the level from the sensor's last update, $\tilde{b}_k(t)$.
	
	Due to the partial knowledge of the battery levels at the edge node, it may happen that the edge node commands sensor $k$ to send a status update (i.e., $a_k(t) = 1$), while sensor $k$ has run out of battery (i.e., $b_k(t) = 0$). Consequently, the sensor can not send a status update (i.e., $d_k(t) = 0$). In this case, the edge node does not receive any status update from the sensor during slot $t$, and thus, serves the user's request using the previous measurement from the local cache. 
	
	In conclusion, sensor $k$ sends a status update packet only whenever it is commanded by the edge node and it has at least one unit of energy in its battery, i.e., 
	\begin{equation}\label{constraint2}
	d_k(t) =  a_k(t) \mathds{1}_{\{b_k(t) > 0\}},
	\end{equation}
	where $\mathds{1}_{\{.\}}$ is the indicator function.
	
	\subsection{Age of Information}
	\textit{Age of information} (AoI) is a \textcolor{black}{destination-centric} metric that quantifies the freshness of information of a remotely observed random process \cite{AoI_Orginal_12,yates19AoI,costa16AoI}. Formally, AoI is the time elapsed since the
	generation of the last received status update packet.
	\textcolor{black}{Let $\Delta_k(t)$ be the \textit{age} of the value of $f_k$ at the edge node at the beginning of slot  $t$, i.e., the number of slots elapsed since the generation timestamp of the last received status update packet from sensor $k$. More precisely, $\Delta_k(t) = t - u_k(t)$ where $u_k(t)$ represents the most recent time slot in which the edge node received a status update packet from sensor $k$, i.e., $u_k(t) = \max \{t'| t'<t, d_k(t') = 1 \}$.}
	Accordingly, the evolution of  $\Delta_k(t)$ can be written as
	\begin{equation}\label{AoI}
	\Delta_k(t+1)=
	\begin{cases}
	\Delta_k(t)+1,&\text{if}~d_k(t)=0\\
	1,&\text{if}~d_k(t)=1,
	\end{cases}
	\end{equation}
	which can be expressed compactly as  $\Delta_k(t+1)= \left(1-d_k(t) \right) \Delta_k(t)+1$.

	\subsection{Cost Function and Problem Formulation}\label{sec_cost}
	We consider a cost function that has two components: one penalizes the energy consumption (characterized by $d_k(t)$) and the other one penalizes the information staleness. 
	More precisely, we define the cost of serving a request for the value of  physical quantity $f_k$ at slot $t$ (i.e., $r_k(t) = 1$)  as 
	\begin{equation}\label{cost_fcn}
	c_k(t) =  (1-\beta) d_k(t)+ \beta r_k(t) g_k(\Delta_k(t+1)),
	\end{equation}
	where a weighting parameter $\beta\in[0,1]$ determines the trade-off between the emphasis on energy consumption and information staleness, and $g_k(\cdot)$ is an increasing function of AoI (see e.g. \cite{kosta2017nonlinearage,zheng2019closedFormNonlinearAge,Sun2019NonlinearAge}). 
	
	We aim to find  the best action of the edge node at each time slot, which is called an \textit{optimal policy}, that minimizes the time-average accumulated cost, defined as
	\begin{equation}\label{average_cost}
	\bar{C}=\displaystyle\underset{T\rightarrow\infty}{\mathrm{lim}}\frac{1}{T}\displaystyle\sum_{t=1}^{T}\displaystyle\sum_{k=1}^{K} c_k(t).
	\end{equation}
	\textcolor{black}{The cost in \eqref{average_cost} can be equivalently expressed as
		\begin{equation}
		{\bar{C} = \sum_{k=1}^{K}\bar{C}_k},
		\end{equation}
		where  $\bar{C}_k$ is the time-average accumulated cost associated with sensor $k$, defined as
		\begin{equation}\label{average_cost_persensor}
		\bar{C}_k=\displaystyle\underset{T\rightarrow\infty}{\mathrm{lim}}\frac{1}{T}\displaystyle\sum_{t=1}^{T} c_k(t),~k = 1, \dots, K.
		\end{equation}}
	\textcolor{black}{\begin{remark}\label{rem1} 
			\normalfont{Focusing on finding $a_k(t)$, $k \in \mathcal{K}$, that minimizes \eqref{average_cost}, we conclude that the above problem is  \textit{separable} across $k$. Namely, the decisions of the edge node for each sensor do not affect the decisions for the others, i.e., the actions $a_k(t)$ are independent across $k\in\mathcal{K}$.}
	\end{remark}}
	
	\textcolor{black}{By Remark \ref{rem1}, minimizing the system-wise cost in \eqref{average_cost} reduces to minimizing the $K$ per-sensor time-average accumulated costs in \eqref{average_cost_persensor}. This will be a key factor for developing our algorithm in  Section \ref{sec:rl}.}	
	\begin{remark}
		\normalfont{Note that in searching for the policy that minimizes \eqref{average_cost_persensor}, only the selection of those actions $a_k(t)$ for which $r_k(t)=1$ \textcolor{black}{needs to} be optimized. Namely, it is clear that if $r_k(t) = 0$, the best action is $a_k(t) = 0$; this implies $d_k(t)=0$, and consequently,   $c_k(t)=0$.}
	\end{remark}
	
	In the next section, we model the problem of minimizing the average cost over all sensors in \eqref{average_cost} (which is equal to minimizing $K$ per-sensor average costs in \eqref{average_cost_persensor}) as a Markov decision process (MDP) and search for the optimal policy using reinforcement learning (RL) \cite{sutton2018reinforcement}.

	\section{Reinforcement Learning Based Status Update Policy}\label{sec:rl}
	\textcolor{black}{In this section, we model the problem of finding an optimal policy at the edge node as an MDP and propose a Q-learning based algorithm to find an optimal policy that minimizes the expected long-term cost. As a key advantage, the proposed algorithm is simple with low complexity of implementation, which is an important point in practice.}
	
	\subsection{MDP Modeling}
	The MDP model can be defined by the tuple ${\left\lbrace \mathcal{S}, \mathcal{A}, \mathcal{P} \left( s(t+1) |s(t), a(t)\right) , c(t), \gamma \right\rbrace}$, where
	\begin{itemize}
		\item $\mathcal{S} = \mathcal{S}_1\times \dots \times \mathcal{S}_K$ is the set of system states, where $\mathcal{S}_k$ is the per-sensor state set. Let $s(t) \in \mathcal{S}$ denote the state at slot $t$, which is equal to $s(t) = \left\lbrace s_1(t), \dots, s_K(t) \right\rbrace $. At each time slot, the per-sensor state $s_k(t) \in \mathcal{S}_k$ is characterized by 1) partial knowledge about the battery level of sensor $k$, i.e., $\tilde{b}_k(t) = b_k(u_k(t))$, and 2) the AoI of the value of $f_k$ in the local cache $\Delta_k(t)$. Thus, $s_k(t) = \left\lbrace \tilde{b}_k(t), \Delta_k(t) \right\rbrace $.
		It is important to point out that the state contains $\tilde{b}_k(t)$ instead of $b_k(t)$, because  the edge node is unaware of the exact battery level of sensor $k$ at slot $t$.
		\item $\mathcal{A}= \mathcal{A}_1\times \dots \times \mathcal{A}_K$ is the  action set, where $\mathcal{A}_k = \left\lbrace 0,1 \right\rbrace $ is the per-sensor action set. The action selected by the edge node at slot $t$ is denoted by $a(t) \in \mathcal{A}$, which is defined as $a(t) = \left\lbrace a_1(t),\dots, a_K(t) \right\rbrace $, $a_k(t) \in \mathcal{A}_k$.
		\item $\mathcal{P} \left( s(t+1) |s(t), a(t)\right)$ is the state transition probability that maps a state-action pair at time slot $t$ onto a distribution	of states at time slot $t + 1$.
		\item $c(t)$ is the immediate cost function, i.e., the cost of taking action $a(t)$ in state $s(t)$, which is defined as $c(t) = \left\lbrace c_1(t), \dots, c_K(t)\right\rbrace $.
		\item $\gamma \in \left( 0 , 1\right]  $ is the discount factor used to weight the immediate cost relative to the future costs. In general, the factor $\gamma$ is smaller than one to guarantee that the cumulative reward is finite, given that the immediate cost is bounded \cite{sutton2018reinforcement}.
	\end{itemize}
	
	The long-term accumulated cost  is defined as 
	\begin{equation}
	C(t) =  \displaystyle\sum_{k = 1}^{K} C_k(t), 
	\end{equation}
	where $ {C_k(t) = \textstyle\sum_{\tau = 0}^{\infty} \gamma ^ \tau c_k(\tau + t)}$.
	Formally, policy $\pi = \pi(a(t)|s(t))$ is defined as a mapping from state $s(t)$ to a probability of choosing action $a(t)$. Note that $\pi = \left\lbrace \pi_1, \dots, \pi_K\right\rbrace $, where $\pi_k = \pi_k(a_k(t)|s_k(t))$, $k\in \mathcal{K}$.  
	Our optimization problem is to find an optimal policy that minimizes the expected long-term accumulated cost over all sensors, i.e.,  $\pi^* = \argmin_{\pi} \mathbb{E_{\pi}}\left[C(t) \mid \pi \right]$. According to Remark \ref{rem1}, the  optimization problem is separable across $k$, and thus, $\pi^* = \left\lbrace \pi^*_1, \dots, \pi^*_K\right\rbrace $ can be found by solving $K$ sub-problems 
	\begin{equation}
	\pi^*_k = \argmin_{\pi_k} \mathbb{E}_{\pi_k}\left[C_k(t) \mid \pi_k \right],~k\in\mathcal{K}.
	\end{equation}

	The state-value and action-value functions are defined to evaluate a policy $\pi$. The state-value function of a state $s$ under a policy $\pi$, denoted by $v_{\pi} \left( s \right)$, is the expected return when starting in state $s$ and following the policy $\pi$ thereafter, i.e.,
	$v_{\pi} \left(s \right)  \doteq \mathbb{E}_{\pi} \left[ C(t)|s(t) = s \right] , \forall s\in \mathcal{S}$.
	The action-value function,  denoted by ${q_{\pi}\left( s,a \right)}$, \textcolor{black}{is the expected return for taking an action $a$ in state $s$ and thereafter following the policy $\pi$}, i.e.,
	$q_{\pi} \left( s,a \right) \doteq \mathbb{E}_{\pi} \left[ C(t)|s(t) = s, a(t) = a \right] , \forall s\in \mathcal{S} , a\in \mathcal{A}$.
	
	The optimal action-value function for state $s$ and action $a$ is defined as ${q^* \left( s,a \right) \doteq \min_{\pi} q_{\pi} \left( s,a \right)}$.
	If  $q^* \left( s,a \right)$ is available, the optimal policy  $\pi^*$ is obtained simply by choosing the action $a$ that minimizes $q^* \left( s,a \right)$ in each state. \textcolor{black}{By using Remark \ref{rem1}, we have  $q^* \left( s,a \right) = \sum_{k=1}^{K} q^*_k(s_k(t),a_k(t))$, where  ${q^*_k \left( s,a \right) = \min_{\pi_k} q_{\pi_k} \left( s,a \right)}$ and $q_{\pi_k} \left( s,a \right) = \mathbb{E}_{\pi_k} \left[ C_k(t)|s_k(t) =s , a_k(t)=a\right]$.}
	
	If the state transition probabilities $\mathcal{P} \left( s(t+1)|s(t), a(t)\right)$, $s\in\mathcal S$, $a\in\mathcal A$,   are available, the optimal policy can be found by dynamic programming, e.g., \textcolor{black}{by the model-based methods such as the value iteration algorithm} \cite[Ch. ~4]{sutton2018reinforcement}. Since $\mathcal{P} \left( s(t+1)|s(t), a(t)\right)$ is \textit{unknown} in our considered scenario, we use model-free RL to learn the action-value functions \textit{by experience}.

	\subsection{Online Q-learning Algorithm} \label{subsec:algorithm}
	Q-learning is an \textit{online} model-free RL algorithm that finds the optimal policy iteratively. In the Q-learning method, the learned action-value function for sensor $k$, denoted as $Q_k$, $k \in \mathcal{K}$, directly approximates the optimal action-value function $q^*_k(s,a)$, $\forall s\in \mathcal{S}_k,a \in \mathcal{A}_k$
	\cite[Sect.~6.5]{sutton2018reinforcement}. The convergence $Q_k \rightarrow q^*_k$ requires that all state-action pairs continue to be updated. To satisfy this condition, a typical approach is to use the "exploration-exploitation" technique in the action selection. The $\epsilon$-greedy algorithm is one such method that trade-offs exploration and exploitation \cite[Sect.~6.5]{sutton2018reinforcement}.
	
	Our proposed Q-learning algorithm is presented in Algorithm \ref{q-algorithm}. 
	To allow exploration-exploitation, the edge node takes either a random or greedy action at slot $t$; the probability of taking a random action is denoted by $ \epsilon(t) $, and thus, the probability of exploiting the greedy action $a_k(t) = \argmin_{a\in \mathcal{A}_k} Q_k(s_k(t), a)$ is $1-\epsilon(t)$.  Generally, during initial iterations,  it is better to set $\epsilon(t)$ high in order to learn the underlying dynamics, i.e., to allow more exploration. On the other hand, in stationary settings and once enough observations are made, small values of $\epsilon(t)$ become preferable to increase tendency to exploitation.
	
	\begin{algorithm}[t]
		\caption{\textcolor{black}{Status update control algorithm via  Q-learning}\label{q-algorithm}}
		\begin{algorithmic}[1]
			\STATE \textbf{Initialize} $Q_k(s,a) = 0$, $\forall s \in \mathcal{S}_k, a \in \mathcal{A}_k$, $k \in \mathcal{K}$
			\FOR{each slot $t = 1, 2, 3, \dots$}
			\FOR{$k = 1, \dots, K$}
			\IF {$r_k(t)=0$} 
			\STATE  $a_k(t) = 0$ 
			\ELSE
			\STATE $a_k(t)$ is chosen according to the following probability\\
			$ a_k(t)  =\left\{
			\begin{array}{ll}
			\argmin_{a\in \mathcal{A}_k} Q(s_k(t), a) \hspace{0mm},\text{w.p.} \hspace{0mm}  1-\epsilon(t)\\
			\textrm{a random action } a\in \mathcal{A}_k \hspace{0mm},\text{w.p.} \hspace{0mm} \hspace{5mm} \epsilon(t)
			\end{array}
			\right.$
			\IF{$a_k(t) = 1$} \STATE Command sensor $k$ to send a status update packet
			\IF {$b_k(t)>0$} \STATE  $d_k(t) = 1$
			\ELSE \STATE $d_k(t) = 0$
			\ENDIF
			\ELSE \STATE $d_k(t) = 0$
			\ENDIF
			\ENDIF
			\STATE Update AoI according to (3) and calculate $c_k(t)$
			\ENDFOR
			\STATE Wait for the next requests and compute $s(t+1)$
			\FOR[\textcolor{black}{Update Q-tables}]{each sensor $k = 1, \dots, K$} 
			\STATE $\begin{array}{ll} Q_k\left( s_k(t),a_k(t)\right) \gets  (1-\alpha(t))Q_k\left( s_k(t),a_k(t)\right) \\ \hspace{8mm}+ \alpha(t) \left(c_k(t)+\gamma \min_{a\in\mathcal{A}_k} Q_k\left(s_k(t+1),a \right)  \right) \end{array}$
			\ENDFOR
			\ENDFOR
		\end{algorithmic}
	\end{algorithm}
	\section{Simulation Results}\label{sec_simulation}
	In this section, simulation results are presented to demonstrate the benefits of the proposed Q-learning method summarized in Algorithm \ref{q-algorithm}.

	\begin{figure}[t]
		\centering
		\includegraphics[width=1\columnwidth]{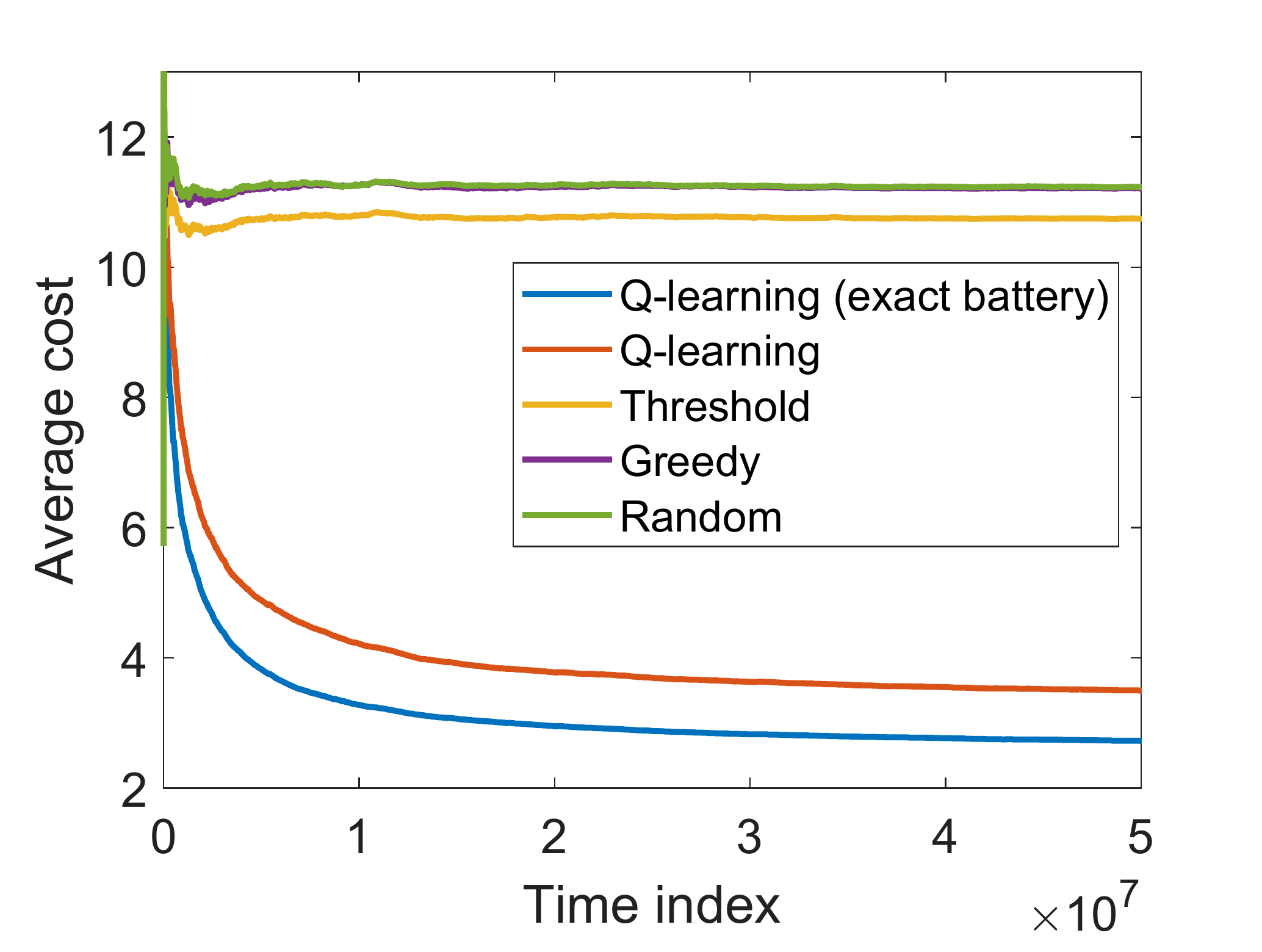}
		\caption{Convergence behavior of the proposed algorithm and baseline methods for weighting parameter $\beta= 0.6$.}
		\label{fig_learningcurve}
	\end{figure}
	
	\begin{figure}[t]
		\begin{subfigure}{1 \columnwidth}
			\centering
			\includegraphics[width=.9 \columnwidth]{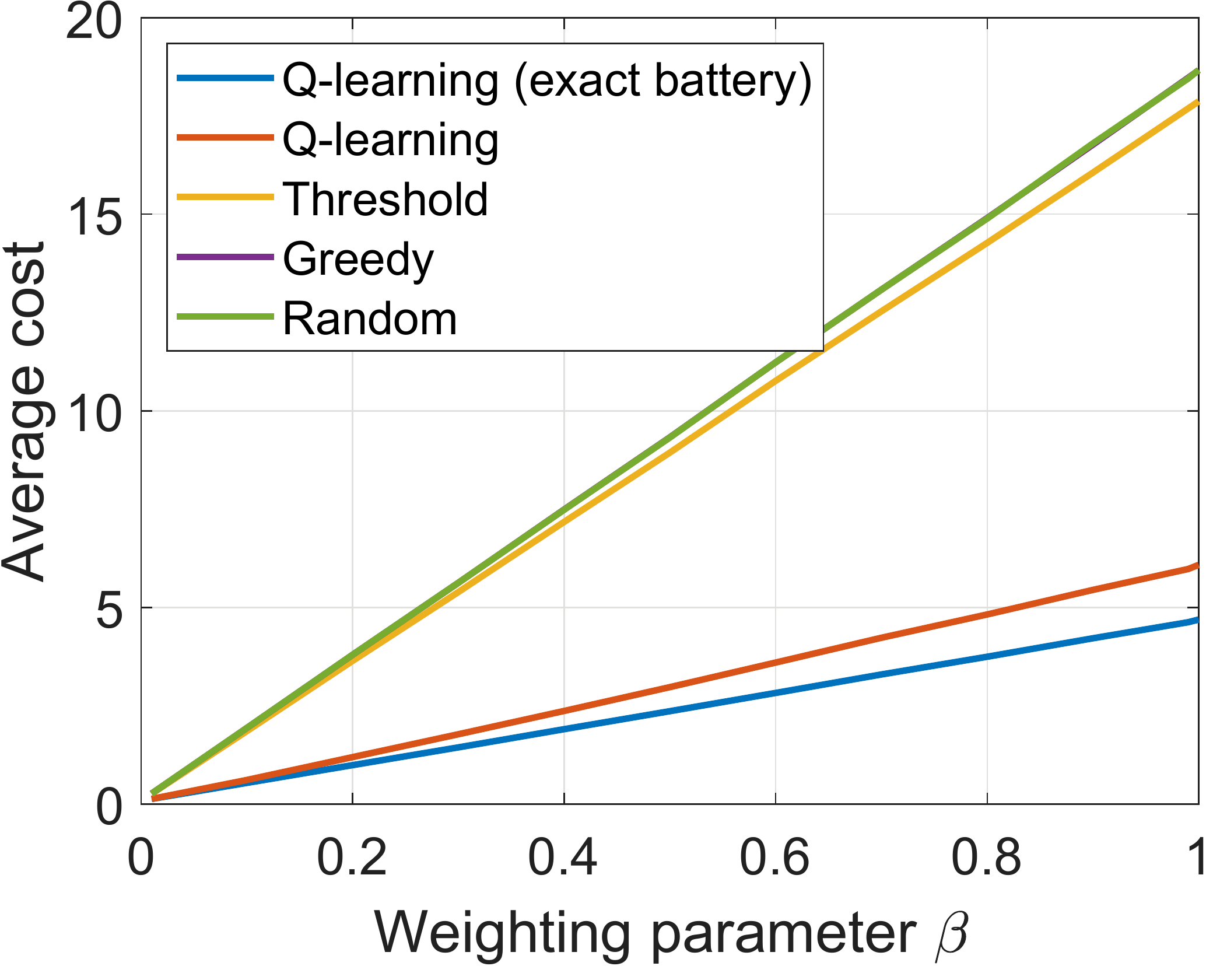}  
			\caption{}
			\label{fig_wvec_a}
		\end{subfigure}
		\begin{subfigure}{1 \columnwidth}
			\centering
			\includegraphics[width=.9 \columnwidth]{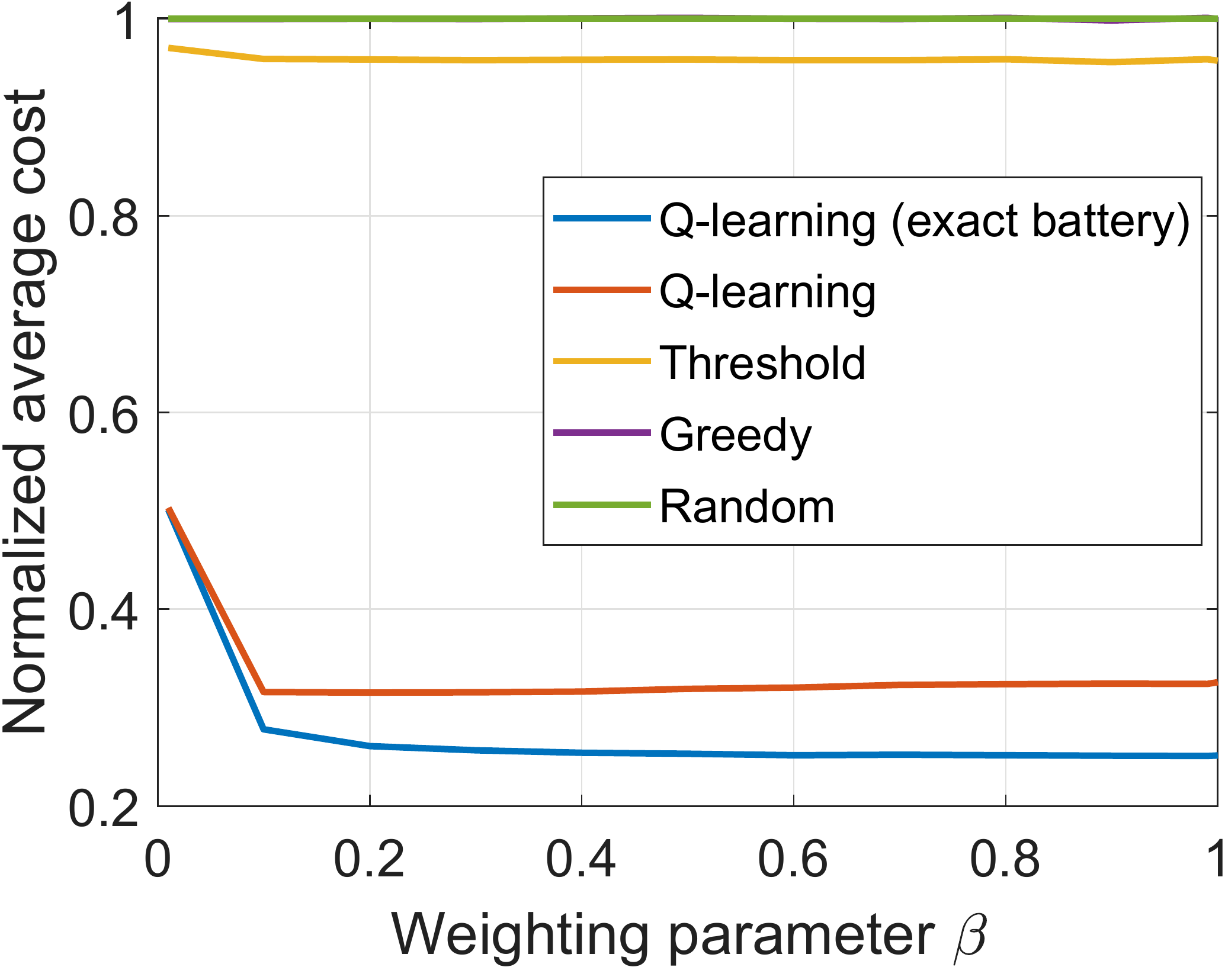}  
			\caption{}
			\label{fig_wvec_b}
		\end{subfigure}
		\caption{Performance of our Q-learning algorithm and other baseline methods as a function of weighting parameter $\beta$ in terms of (a) average cost and (b) normalized average cost.}
		\label{fig_wvec}
	\end{figure}

	\subsection{Simulation Setup}
	\textcolor{black}{The simulation scenario consists of $K = 3$ energy harvesting sensors, i.e.,  $\mathcal{K}=\left\lbrace 1,2,3 \right\rbrace$.} Each sensor has a battery of finite capacity $B = 10$ units of energy. 
	At each time slot the probability that the value of $f_k$ is requested (i.e., $r_k(t) = 1$)  is denoted by $p_k$, i.e., $\mathrm{Pr}\{r_k(t) = 1\} = p_k$.
	We set $p_k=0.1$.
	
	We model the underlying energy harvesting process of sensor $k$  as a two-state Markov chain with state space  $\left\lbrace V_{k,1},V_{k,2}\right\rbrace$. For example, the states can represent "good" and "bad" energy harvesting states \cite{michelusi2013transmission}. Let $V_k(t)$ denote the state of the environment at slot $t$ for sensor $k$.  At slot $t$, if $V_k(t) = V_{k,1}$,  sensor $k$ harvests one unit of energy (i.e., $e_k(t) = 1$) with probability $\lambda_{k,1}$, i.e., $\mathrm{Pr}(e_k(t) =1|V_k(t)=V_{k,1}) =  \lambda_{k,1}$. Similarly, if $V_k(t) = V_{k,2}$, sensor $k$ harvests one unit of energy with probability $\lambda_{k,2}$. We denote the transition probability from state $V_{k,i}$ to state $V_{k,j}$ by $p_{k,ij} = \mathrm{Pr}(V_k(t) = V_{k,i}|V_k(t-1) = V_{k,j})$, $i,j\in\{1,2\}$. We set $\lambda_{k,1} = 0.04$, $\lambda_{k,2} = 0.0004$, $p_{k,11} = 0.7$, $p_{k,12} = 0.3$, $p_{k,21} = 0.6$, and  $p_{k,22} = 0.4$, $k \in \mathcal{K}$\footnote{In  general, one can consider different energy harvesting models among the sensors for the proposed method, i.e., different values for $p_{k,ij}$, $\lambda_{k,1}$, and  $\lambda_{k,2}$ for each $k \in \mathcal{K}$.}.
	
	For determining the cost function in \eqref{cost_fcn}, we define the function $g_k(\Delta_k(t+1))$  as
	\begin{equation}\label{freshness_loss}
	g_k(\Delta_k(t+1))= {\left(\frac{\Delta_k(t+1)}{\zeta_k}\right)}^{\mu},
	\end{equation} 
	where ${\zeta_k}$ is the tolerance of using aged measurements of $f_k$, and  ${\mu \ge1}$ is a parameter that adjusts how aggressively we penalize when the AoI has a higher value than the tolerance of $f_k$, i.e., when ${\Delta_k(t+1)>\zeta_k}$. The function in \eqref{freshness_loss} is  a scaled version of a non-linear AoI; different functions for non-linear AoI have been investigated in \cite{kosta2017nonlinearage,zheng2019closedFormNonlinearAge,Sun2019NonlinearAge}. Note that with $\mu=1$ and $\zeta_k=1$, $\forall{k}\in \mathcal{K}$, \eqref{freshness_loss} is purely characterized by the AoI, i.e., $g(\Delta_k(t+1))=\Delta_k(t+1)$.  We set $\mu = 2$ and  select $\zeta_k$ uniformly random from the interval $[3 \ 15]$. \textcolor{black}{Note that other functions are also applicable, e.g., $g(\Delta_k(t+1))=\textrm{log}(1+\Delta_k(t+1))$ \cite{kosta2017nonlinearage}.}
	
	In Algorithm \ref{q-algorithm}, we set $\epsilon(t)  =0.02+ 0.98 e ^{-\epsilon_\textrm{d}t}$ with decay parameter $\epsilon_\textrm{d} = 0.01$, and the discount factor as $\gamma = 0.99$. The learning rate $\alpha(t)$ is set to $\alpha(t) = 0.5$ during the first $1/\epsilon_\textrm{d}=100$ iterations and  after that $\alpha(t) = 0.1$.
	
	We evaluate the performance of the proposed algorithm in terms of average cost defined in \eqref{average_cost}. Three baseline policies are considered: \textit{greedy}, \textit{threshold}, and  \textit{random}.    
	In the greedy policy,  whenever the value of $f_k$ is requested (i.e., $r_k(t)=1$), the edge node commands sensor $k$ to send a status update (i.e., $a_k(t)=1$); sensor $k$ sends a status update if the battery is non-empty, $b_k(t) > 0$.
	In the threshold policy, whenever the value of $f_k$ is requested (i.e., $r_k(t)=1$) and $ \Delta_k(t)+1 >\zeta_k$, the edge node commands sensor $k$ to send a status update. \textcolor{black}{In the random policy, a random action $a_k(t)\in \{0,1\}$ is selected at each time slot}. 
	For the benchmarking, we also consider a \textit{genie-aided} Q-learning method that knows the exact battery level of all sensors at each time slot. This policy serves clearly as a lower bound to the proposed Q-learning algorithm. 
	
	\subsection{Results}
	Fig.~\ref{fig_learningcurve} depicts the learning curves of each algorithm for weighting parameter ${\beta= 0.6}$. The proposed Q-learning algorithm significantly outperforms other baseline methods; the decrease of the average cost is roughly threefold compared to the threshold algorithm, which has the best performance among the baseline policies here.
	Interestingly, the gap between the proposed Q-learning algorithm and  the genie-aided Q-learning algorithm is small. This demonstrates that the proposed algorithm has high performance even it only has the partial knowledge about the battery levels of the sensors at each time slot, which is the case in practice.


	Next, we focus on the average cost obtained by averaging each algorithm over $5$ episodes where each episode takes $3\times10^7$ iterations. Fig.~\ref{fig_wvec}(a) illustrates the  average cost of each algorithm for different values of weighting parameter $\beta$. For better visualization, Fig.~\ref{fig_wvec}(b) depicts a normalized average cost of each algorithm, defined as the ratio of the average cost of each algorithm to the average cost of the random policy.   
	As shown in Fig.~\ref{fig_wvec}(a), when $\beta$ increases, the average cost increases for all algorithms, because the second term of  \eqref{cost_fcn} is squared ($\mu = 2$) (see \eqref{freshness_loss}). 
	\textcolor{black}{As shown in Fig.~\ref{fig_wvec}(b), for all values of $\beta$, the proposed Q-learning algorithm, which does not know the exact battery levels, 
		performs close to the genie-aided Q-learning algorithm. Furthermore, the Q-learning algorithm reduces the average cost approximately by a factor of 3 compared to the threshold algorithm.}
	
	
	
	\section{Conclusions}
	We investigated a status update control problem  in an IoT sensing network consisting of multiple users, multiple energy harvesting sensors, and a wireless edge node. We modeled the problem as an MDP and proposed an RL based algorithm that finds an optimal status update control policy that  minimizes the average long-term cost which strikes a balance between the AoI and energy consumption. The proposed scheme does not need any information about the energy harvesting model and the exact battery level of the sensors. Simulation results showed  the advantage of the proposed Q-learning algorithm.
	
	\section{Acknowledgments}
	This research has been financially supported by the Infotech Oulu, the Academy of Finland (grant 323698), and Academy of Finland 6Genesis Flagship (grant 318927). The work of M. Leinonen has also been financially supported in part by the Academy of Finland (grant 319485).
	M. Codreanu would like to acknowledge the support of the European Union's Horizon 2020 research and innovation programme under the Marie Sk\l{}odowska-Curie Grant Agreement No. 793402 (COMPRESS NETS).

	\bibliographystyle{IEEEtran}
	\bibliography{conf_short,IEEEabrv,Bibliography}
	
\end{document}